\documentclass[prd,aps,twocolumn,amsfonts,showpacs,superscriptaddress,preprintnumbers,nofootinbib]{revtex4-1}

\usepackage{graphicx}
\usepackage{xcolor}
\usepackage{rotating}
\usepackage{bm,amsmath,amssymb}
\usepackage[utf8]{inputenc}
\usepackage{footmisc}
\usepackage{diagbox}

\usepackage[colorlinks=true, urlcolor=blue, linkcolor=blue, citecolor=blue, hyperindex=true, linktocpage=true]{hyperref}

\newcommand{\nn}{\nonumber\\}

\def\CN{\mathcal{N}}

\def\le{\left}
\def\ri{\right}

\def\be{\begin{equation}}
\def\ee{\end{equation}}

\def\bea{\begin{eqnarray}}
\def\eea{\end{eqnarray}}
\newcommand{\bma}{\le(\begin{matrix}}
\newcommand{\ema}{\end{matrix}\ri)}

\newcommand{\bega}{\begin{gather}}

\begin{document}

\preprint{IFT-UAM/CSIC-23-94}

\title{Reconstruction of the quasinormal spectrum from pole-skipping}

\author{Sa\v{s}o Grozdanov}
\affiliation{Higgs Centre for Theoretical Physics, University of Edinburgh, Edinburgh, EH8 9YL, Scotland, 
}
\affiliation{Faculty of Mathematics and Physics, University of Ljubljana, Jadranska ulica 19, SI-1000 Ljubljana, Slovenia
}

\author{Timotej Lemut}
\affiliation{Faculty of Mathematics and Physics, University of Ljubljana, Jadranska ulica 19, SI-1000 Ljubljana, Slovenia 
}

\author{Juan F. Pedraza}
\affiliation{Instituto de F\'{i}sica T\'{e}orica UAM/CSIC, Calle Nicol\'{a}s Cabrera 13-15, Madrid 28049, Spain
}

\begin{abstract}
The holographic gauge/gravity duality provides an explicit reduction of quantum field theory (QFT) calculations in the semi-classical large-$N$ limit to sets of `gravitational' differential equations whose analysis can reveal all details of the spectra of thermal QFT correlators. We argue that in certain cases, a complete reconstruction of the spectrum and of the corresponding correlator is possible from only the knowledge of an infinite, discrete set of pole-skipping points traversed by a single (hydrodynamic) mode computed in a series expansion in an inverse number of spacetime dimensions. Conceptually, this reduces the computation of a QFT correlator spectrum to performing a set of purely algebraic manipulations. With the help of the pole-skipping analysis, we also uncover a novel structure underpinning the coefficients that enter the hydrodynamic dispersion relations.  
\end{abstract}
\maketitle

\section{Introduction}
Quantum field theories (QFTs) typically exhibit intricate and complicated dynamical properties that result from the interactions between their degrees of freedom. While no all-encompassing definition of a QFT exists, one way to think of them is through the spectral properties of correlation functions of {\it all} operators. Given that such computations can be extremely difficult and often even outside the scope of our methodological and technical capabilities, a natural question to ask is, therefore, what is the minimal amount of information that completely determines the spectrum of each such correlator?

The advent of the gauge/gravity (holographic) duality \cite{Maldacena:1997re} introduced a new way of looking at QFTs through the lens of gravity. In its simplest form, it taught us that certain strongly coupled QFTs with a large local number of degrees of freedom (large-$N$ theories) are dual to classical field theories that include gravity (general relativity). 

It has long been appreciated that large-$N$ QFTs are semi-classical and that the $N\to\infty$ limit can be understood as an effective $\hbar \to 0$ limit (see~Ref.~\cite{RevModPhys.54.407}). Nevertheless, in general, solving such theories by computing various correlation functions is essentially impossible. In principle, this could be done if one were for example able to sum an infinite number of planar Feynman diagrams \cite{tHooft:1973alw}, solve the master field equations \cite{Witten:1979kh} or solve the problem by using the method of coadjoint orbits \cite{RevModPhys.54.407}. Nevertheless, rather spectacularly, in holographic $N \to \infty$ theories, one can use the duality to turn such problems into solving partial differential equations that arise as classical equations of motion of the dual gravitational field theory. Therefore, holography dramatically reduces the difficulty in computing (at least strongly-coupled) correlation functions in certain $N \to \infty$ QFTs.

In this Letter, we will argue that through the phenomenon of pole-skipping \cite{Grozdanov:2017ajz,Blake:2017ris,Blake:2018leo,Grozdanov:2018kkt} and the introduction of another perturbative parameter, which will be essential for deriving our results: $1/d$, where $d$ is the number of spacetime dimensions of the gauge theory (see Ref.~\cite{Emparan:2020inr}), one can turn the problem of computing certain holographic QFT correlation functions into a purely local and algebraic problem involving an expansion around the dual event horizon with no need for solving differential equations in the entire bulk. Even if these algebraic problems remain extremely hard to solve, we find this to be a conceptually important step towards the answer to the following question: ``How much information is required to determine the spectrum of a QFT correlation function?''

\section{The claim}
Consider a neutral CFT${}_d$ at non-zero temperature $T$ in $d$ spacetime dimensions with a classical holographic dual. In $d=4$, this can be the infinitely strongly coupled $\CN = 4$ supersymmetric Yang-Mills (SYM) theory with an infinite number of colours. We will study, for concreteness, the spectra of transverse momentum two-point functions. In each such holographic CFT${}_d$, the spectrum equals that of the dual transverse gravitational quasinormal modes (QNMs) in the background of the Schwarzschild-AdS${}_{d+1}$ black brane in Einstein gravity \cite{Kovtun:2004de}. These QNMs follow from the differential equation describing the diffeomorphism-invariant transverse bulk mode $Z(r)$ in any $d\geq 3$ (see e.g.~Ref.~\cite{Morgan:2009pn}): 
\begin{align}
&\partial_r \left[ \frac{r^{d+1}\left(r^2 f(r) \partial_r Z(r) - i \omega Z(r) \right) }{\omega^2 r^2 - q^2 r^2 f(r)} \right] - \nn
&- \frac{r^{d-1} \left(i \omega r^2 \partial_r Z(r) + q^2 Z(r) \right)}{\omega^2 r^2 - q^2 r^2 f(r)} = 0, \label{EoM}
\end{align}
where we have used the ingoing Eddington-Finkelstein coordinates, $f(r) = 1 - r^{-d}$, the event horizon $r_h$ is placed at $r_h = 1$ and its temperature is $T = d/4\pi$. Moreover, $\omega$ and $q = |{\bf q}|$ are the frequency and the momentum (wavevector). 

Such spectra contain an infinite set of modes $\omega_i$, where $i = 0,1,\ldots$. Among them is $\omega_0$, which is a gapless, (transverse) diffusive hydrodynamic mode whose dispersion relation has the following series representation:
\begin{equation}\label{Ser_Diffusion}
\omega_0(q) = -i \sum_{n=1}^\infty a_n q^{2n}.
\end{equation}
The series has a finite radius of convergence $R$ \cite{Grozdanov:2019kge,Grozdanov:2019uhi} (see also \cite{Withers:2018srf,Heller:2020hnq}). For real and positive $q$ analytically continued outside $|q| < R$, this dispersion relation passes through an infinite set of pole-skipping points at a sequence of $q_n\in\mathbb{R}$, such that (see Refs.~\cite{Grozdanov:2019uhi,Blake:2019otz})
\begin{equation}\label{PS-w}
\omega_0(q_n) = \omega_n = - 2 \pi T i n, \quad n = 0,1,\ldots.
\end{equation} 
In $d=3$ (CFT${}_3$ dual to a stack of M2 branes), the sequence of $q_n$ is known analytically (see Refs.~\cite{Grozdanov:2020koi,Grozdanov:2023txs}). It is important to note that in the theories considered here, the diffusive mode does not `collide' with any other mode for real positive $q$ (and negative imaginary $\omega$).

For (at least) the spectra in these CFT${}_d$'s, we claim that the knowledge of all pole-skipping points $\omega_0(q_n)$ traversed by the hydrodynamic mode for $q\in\mathbb{R}_+$ (cf.~Eq.~\eqref{PS-w}) is sufficient for reconstructing the entire spectrum of modes and their dispersion relations $\omega_{i\geq 0} (q)$. Since the Green's function is meromorphic in $\omega$, the knowledge of all $\omega_i(q)$ allows us to determine the full denominator of the correlator. Up to a rescaling, the numerator of the correlator then follows from the statements made in the recent work~\cite{Dodelson:2023vrw}.

To substantiate our claim, we will first argue that $\omega_0(q)$ indeed follows from $\omega_0(q_n)$. This is an interpolation problem: we wish to {\it uniquely} determine a function by knowing its values (the values of the analytically continued \eqref{Ser_Diffusion}) at an infinite number of points \eqref{PS-w}. In general, such problems are very difficult and, depending on the complex analytic properties of $\omega(q)$, may or may not have a unique solution.\footnote{An example of an interpolation problem with a unique solution is the Nevanlinna-Pick interpolation in cases with a vanishing determinant of the Pick matrix.} In this case, too, the complicated multi-sheeted Riemann surface structure of $\omega_0(q)$ (see \cite{Grozdanov:2019uhi,Grozdanov:2022npo}) makes this prohibitively difficult in any specific $d$ (even in $d=3$). Nevertheless, we will argue that this problem can in fact be solved by continuing the calculation from a specific $d$ to one in general $d$. Then, the interpolation can be uniquely solved by employing a perturbative $1/d$ expansion around $d\to\infty$, the result resummed and evaluated at a desired $d$. 

The properties of general relativity in the large-$d$ limit are key to this result. They follow from a simplification of the gravitational interaction in such a limit, as can be seen from the sharp
decay of the gravitational potential around a point-like mass, which at large $d$ scales like $1/r^{d}$ with distance $r$. The gravitational field therefore becomes strongly localised near its source \cite{Emparan:2013moa}. With regard to black hole perturbations, an important consequence is the existence of two sets of modes with frequencies parametrically separated in $1/d$. One uncovers a set of (non-decoupled) fast modes with $\omega \sim d/r_h$, and a set of (decoupled) slow modes with $\omega \sim 1/r_h$. The latter ones are completely localised within the near-horizon region at all orders in $1/d$, and describe the entire hydrodynamic regime of the dual holographic QFT \cite{Emparan:2014aba,Emparan:2015rva}.

The remaining steps in the reconstruction of the spectrum and the associated Green's function, which amount to finding all remaining $\omega_{i\geq 1} (q)$, then follow from the arguments and algorithms presented in Refs.~\cite{Withers:2018srf,Grozdanov:2022npo,Dodelson:2023vrw}. 

\section{From pole-skipping to the hydrodynamic dispersion relation and the full spectrum}
\subsection{General interpolation steps for obtaining the hydrodynamic mode}

While this is certainly a non-trivial task in a QFT, the continuation of a problem to an arbitrary number of dimension $d$ in quantum mechanics, followed by an analysis expanding the wavefunction and the energy levels around $d\to\infty$, is a well-known and highly effective method \cite{Witten1980,1990PhR...186..249C}, analogous to the large-$N$ expansion. To facilitate our discussion, what is more important is that the calculation of a CFT correlation function in general $d$ is also particularly suited to a holographic treatment. Concretely, we `continue' the Einstein gravity problem from a set number of bulk dimensions to a `deformed' problem in a theory with the same action, but in $d+1$ bulk dimensions. It is also most convenient to choose the higher-dimensional background solution to have  `compatible' symmetries and macroscopic properties (here, an uncharged black brane). For transverse gravitational perturbations, these two steps lead to Eq.~\eqref{EoM}.

The analysis then proceeds by writing $\omega_0(q)$ as a double expansion: a gradient expansion in powers of $q^2$ and a series in inverse powers of the dimension $1/d$. By analysing Eq.~\eqref{EoM}, we find that the diffusive series \eqref{Ser_Diffusion} takes the following form:
\begin{equation}\label{hydro_series}
\omega_0(q) = - i \bar q^2 - i \sum_{m=2}^\infty \frac{1}{d^m} \sum_{j=2}^m c_{m,j} \bar q^{2j} ,
\end{equation}
where $\bar{q} \equiv q/\sqrt{d}$. The coefficient of the $\bar q^2$ term is set by the `universal' shear viscosity to entropy density ratio $\eta/s = 1/4\pi$ \cite{Kovtun:2003wp,Kovtun:2004de}. The remaining coefficients $c_{m,j}$ depend on the parameters characterising the CFT$_{d}$ and the state (the background). For momentum diffusion in a neutral thermal CFT$_{d}$, analytic closed-form results are known to order $O(1/d^4)$ \cite{Emparan:2015rva}:
\begin{align}
c_{2,2} &= 2\zeta(2) , \label{c2_res}\\
c_{3,2} &= - 4 \zeta(3), \quad c_{3,3} = - 4 \zeta(3), \label{c3_res} \\
c_{4,2} &= 8 \zeta(4), \quad c_{4,3} = 7\times 8\zeta(4),\quad  c_{4,4} = 8 \zeta(4). \label{c4_res}
\end{align}
In fact, by knowing that for general $d$, the first two terms in the dispersion relation are \cite{Natsuume:2008gy,Bhattacharyya:2008mz}
\begin{equation}\label{hydro_ser_q4_resum}
\omega = - i \bar q^2 -i  \frac{H_{2/d}}{d} \bar q^4 + O(\bar q^6) , 
\end{equation} Ref.~\cite{Emparan:2015rva} could use the series representation of the Harmonic number $H_{2/d} = - \sum_{j=1}^\infty  \zeta(j+1) (-2/d)^j$, where $\zeta(m)$ is the Riemann zeta function, to guess that 
$c_{m,2} = (-1)^m 2^{m-1} \zeta(m)$. The fact  that this series in $1/d$ converges for $d > 2$ conforms to the absence of a transverse diffusive channel in $d=2$. The coefficients at higher orders in $\bar q$ are unknown in general $d$ in closed form.\footnote{In $d=4$ (in the $\CN = 4$ SYM theory at infinite coupling), the dispersion relation was computed to order $q^8$ in Ref.~\cite{Grozdanov:2019uhi}. In $d=3$, the methods of Ref.~\cite{Aminov:2023jve} allow for an analytic calculation in terms of multiple polylogarithms to a much higher order.} We derived a complicated integral expression for $c_{m,3}$, which we  numerically evaluate for the first few new coefficients:
\begin{align}
c_{5,3} &\approx -15.814 \times 16 \zeta(5), \label{c53}\\
c_{6,3} &\approx 32.420 \times 32 \zeta(6).
\end{align}

Next, we develop the relation between $c_{m,j}$ and pole-skipping. By analysing the equation of motion near the horizon, e.g., by using the determinant method of Ref.~\cite{Blake:2019otz} (for a closed-form determinant expression, see Appendix~\ref{app:Determinant}), the sequence of pole-skipping points \eqref{PS-w} with $\bar q \in \mathbb{R}$ can also be expressed as a series in $1/d$:
\begin{equation}\label{qbar_def}
    \bar{q}_n = \sqrt{\frac{nd}{2}}\left(1 + \sum_{m=1}^\infty \frac{b_{n,m}}{d^m} \right). 
\end{equation}
Inserting this expression into Eq.~\eqref{hydro_series} then gives an infinite recursive chain of relations between $b_{n,m}$ and $c_{m,j}$. At the first two levels, we have  
\begin{align}
\!\! b_{n,1} &= - \sum_{m=2}^\infty \frac{n^{m-1} c_{m,m}}{2^m}, \label{b1-rec} \\
\!\! b_{n,2} &= - \frac{b_{n,1}^2}{2}  - \sum_{m=2}^\infty \frac{n^{m-1} \left( c_{m+1,m} + 2 m b_{n,1} c_{m,m} \right)}{2^{m}} , \label{b2-rec}
\end{align}
and so on for higher $b_{n,m\geq3}$. 

Our claim is that the knowledge of $b_{n,m}$ (i.e., the pole-skipping momenta expanded around $d\to\infty$) is sufficient to derive all hydrodynamic coefficients $c_{m,j}$. To do this, we must assume that a (unique) analytic continuation of the coefficients $b_{n,m}$ is possible from $n \in \mathbb{Z}_{\geq 0}$ to $x \in \mathbb{R}$ (or $x \in \mathbb{C}$), such that each $b_m(x)$ is analytic at least in the vicinity of $x = 0$ and for $x \geq 0$. Consider first Eq.~\eqref{b1-rec}. By continuing $b_{n,1}\to b_1(x)$, $b_1(x)$ becomes the {\it generating function} for the hydrodynamic coefficients $c_{m,m}$. In particular, $c_{m,m}$ can be computed by taking derivatives of $b_1(x)$ and evaluating them at $x=0$: $c_{2,2} = - 4 \partial_x b_1(0)$, $c_{3,3} = - 4 \partial^2_x b_1(0)$, and, in general, for $m\geq 2$:
\begin{equation}\label{cmm}
c_{m,m} = - \frac{2^m}{(m-1)!} \partial^{m-1}_x b_1(0) .
\end{equation}
Next, we analytically continue $b_{n,2} \to b_2(x)$ and use Eq.~\eqref{b2-rec} as the generating function equation to find all $c_{m+1,m}$. For example, $c_{3,2} = - 4 \partial_x b_2(0)$, $c_{4,3} = - 4 \partial^2_x b_2(0) + 7 c_{2,2}^2/4$, and, in general, for $m\geq 2$:
\begin{align}\label{cm1m}
&c_{m+1,m} = - \frac{2^m \partial^{m-1}_x b_2(0)}{(m-1)!}  \nn
& +  \sum_{j=2}^{m-1} \left(j-\frac{1}{4}\right)  c_{j,j} c_{m-j+1,m-j+1},
\end{align}
having used that $b_1(0) = 0$ (cf.~Eq.~\eqref{b1-rec}) and Eq.~\eqref{cmm}. It is easy to see that in a similar recursive manner, all $c_{m,j}$ from \eqref{hydro_series} follow from a combination of derivatives of the analytically continued $b_m(x)$ evaluated at $x=0$. 

\begin{figure}[ht!] 
\centering
\includegraphics[width=0.45\textwidth]{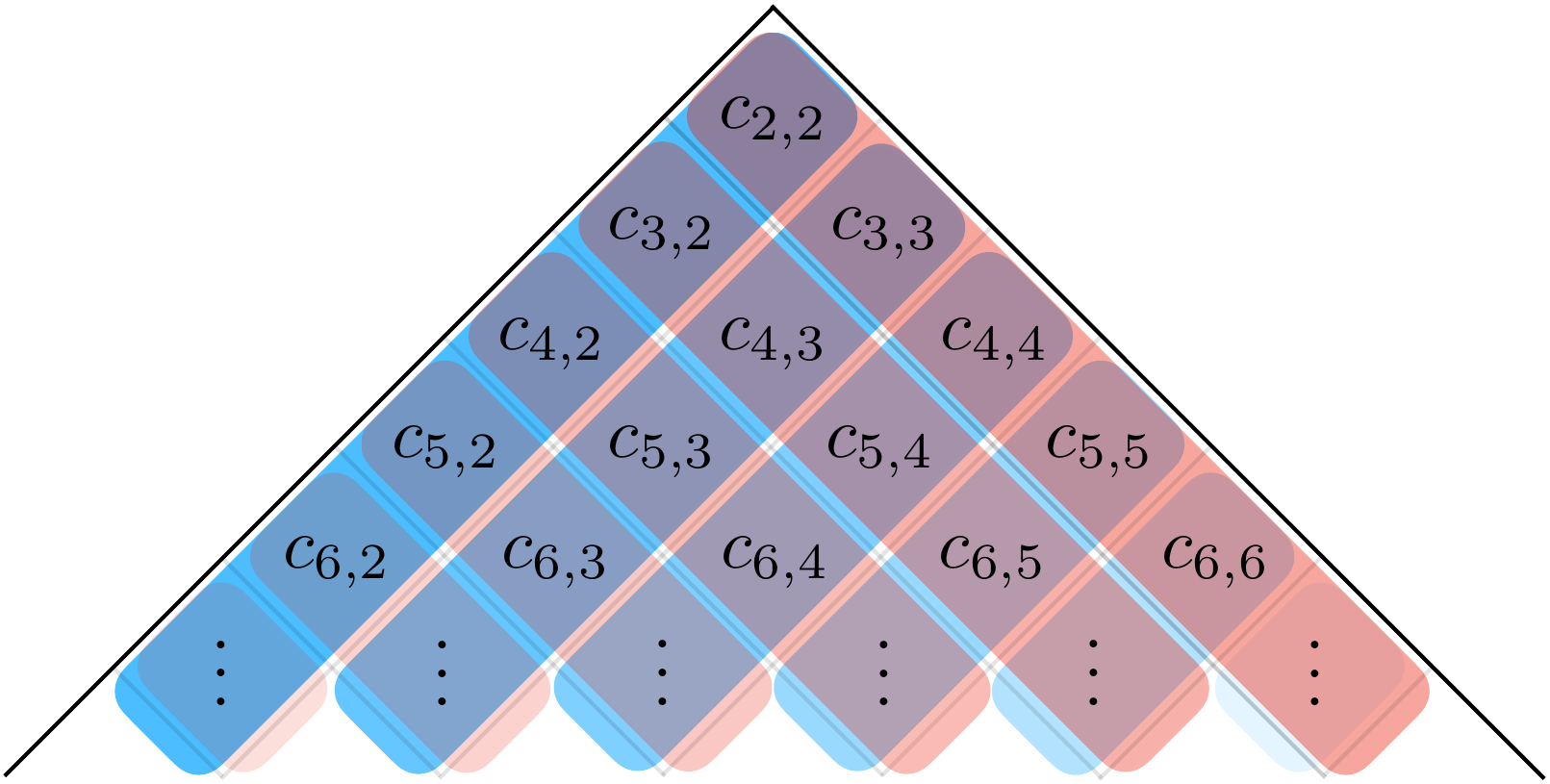}
\caption{Triangular structure depicting the sequences of recovery of the coefficients $c_{m,j}$ in \eqref{hydro_series} when computed from the hydrodynamic gradient expansion (from more to less opaque blue) versus when computed from pole-skipping (from more to less opaque red).}
\label{fig:trikotnik}
\end{figure}

At present, we do not have a general proof that the necessary unique continuation of $b_{n,m} \to b_m(x)$ is always possible. Assuming that in certain cases it is, however, it then follows that the entire hydrodynamic series can be reconstructed from the pole-skipping momenta, which all arise from a local near-horizon analysis. First, $b_{n,1}$ gives $c_{m,m}$. Then, the knowledge of $b_{n,2}$ gives $c_{m+1,m}$, $b_{n,3}$ gives $c_{m+2,m}$, and so on. Recursive pole-skipping calculations thereby reveal a different (`mirrored') set of coefficients as compared to the hydrodynamic gradient expansions, which normally first computes $c_{m,2}$, then $c_{m,3}$, and so on. Our analysis establishes a new (perturbative) approach to accessing a different subset of hydrodynamic coefficients $c_{m,j}$, which we represent in Figure~\ref{fig:trikotnik}.

\subsection{Demonstration of the procedure} 
One way to proceed with the calculation of the pole-skipping $b_{n,m}$ is by using the determinant method of Ref.~\cite{Blake:2019otz}. This procedure is fully systematic, it can be executed in general $d$ and is, moreover, algebraic. We discuss its details in Appendix~\ref{app:Determinant}. The limitation is, however, that the calculation requires setting a specific $n$, which makes it difficult to obtain a closed-form $b_{n,m}$ for all $n\geq 0$, and analytically continue $b_{n,m}\to b_m(x)$. Nevertheless, the method is practical and suitable for a numerical interpolation (`analytic continuation') and calculation of $c_{m,j}$. 

To substantiate the claims made in this work, we may also proceed by computing $b_{n,m}$ from a QNM analysis with a specified $m$ (order in the $1/d$ expansion), but for all $n$. Note, however, that the calculation then requires us again to solve bulk differential equations. We provide the details of this procedure in Appendix~\ref{app:FullB}.

It is easy to find the first-order $b_{n,1}$ for $n\geq 0$ either by using a QNM calculation or the determinant method (and then guessing the result for all $n$). We obtain 
\begin{equation}\label{b1}
    b_{n,1} = - \frac{1}{2} H_n,
\end{equation}
where $H_n = \sum_{k=1}^n 1/k$ is the Harmonic number and $n$ an integer. $H_n$, however, has a well-known analytic continuation to real or complex $x$: $H_n \to H(x) = \sum_{k=1}^\infty x / k(x+k)$. This representation uniquely determines $b_{n,1} \to b_1(x)$ and allows us to compute all $c_{m,m}$ from Eq.~\eqref{cmm}. The result establishes that the outer sides of the hydrodynamic coefficient triangle in Figure~\ref{fig:trikotnik} are mirror images of each other for all $m\geq 2$:
\begin{equation}\label{symmetry_outer}
c_{m,m} = c_{m,2} =  (-1)^m 2^{m-1} \zeta(m).
\end{equation}
With this result, one can perform a partial resummation of the hydrodynamic series~\eqref{hydro_series} along the outer right side of \ref{fig:trikotnik} and obtain
\begin{equation}\label{resum_r1}
    \omega_0(q) = - i \bar q^2 - i \frac{\bar q^2}{d} H_{2 \bar q^2 / d} + \ldots\,.
\end{equation}
The series giving \eqref{resum_r1} converges for $|q^2| < d^2 / 2$. As expected, its convergence properties improve as $d$ increases.

At higher orders, the calculation is significantly more involved. Nevertheless, we could find a closed-form expression for $b_{n,2}$, expressed as a sum, with the help of a QNM calculation discussed in Appendix~\ref{app:FullB}. Here, we use this result to numerically analytically continue $b_{n,2} \to b_2(x)$ using the method of Pad\'{e} approximants to evaluate $\partial^{m-1}_x b_2(0)$ in Eq.~\eqref{cm1m}. In particular, we adjust the order of the approximant to find the best fit for the analytically known coefficients $c_{3,2}$ and $c_{4,3}$ (cf.~Eqs.~\eqref{c3_res}--\eqref{c4_res}). Using $b_{n,2}$ with $n=\{0,1,\ldots,1000\}$ (although significantly fewer coefficients could also be used), the optimal fit is given by the Pad\'{e} approximant of order $[15/13]$.\footnote{The order of the Pad\'{e} approximant $[M/N]$ denotes the rational function form $f(n) = \left(\sum_{i=0}^M a_i n^i\right)/\left(1 + \sum_{j=1}^N b_j n^j\right)$. The fit is performed by using the FindFit command in Mathematica.} Finally, using Eq.~\eqref{cm1m}, the numerical results for the first four $c_{m+1,m}$ are:
\begin{align}
    c_{3,2} &\approx - 1.000 \times 4 \zeta(3), \\
    c_{4,3} &\approx 7.001 \times 8 \zeta(4) , \\
    c_{5,4} &\approx - 15.548 \times 16 \zeta(5), \\
    c_{6,5} &\approx 27.546 \times 32 \zeta(6) .
\end{align}
The estimates of $c_{5,4}$ and $c_{6,5}$, or other (in principle) easily computable higher-order coefficients are new. At this point, however, we cannot ascertain the level of their precision.

Next, we compute $b_{n,3}$ by using the determinant method from Appendix~\ref{app:Determinant} for orders $n=\{0,1,\ldots, 52\}$. The first two coefficients that $b_{n,3}$ generates are $c_{4,2} = - 4 \partial_x b_3(0)$ and $c_{5,3} = - 4 \partial_x^2 b_3(0) + 7 c_{2,2} c_{3,2}/2$. Optimising again the order of the Pad\'{e} approximant by matching to the analytic result $c_{4,2}$ (cf.~Eq.~\eqref{c4_res}), we find the order $[9/13]$, which gives the first two coefficients $c_{m+2,m}$:
\begin{align}
    c_{4,2} &\approx 1.000 \times 8 \zeta(4) ,\\
    c_{5,3} &\approx - 15.502 \times 16 \zeta(5). 
\end{align}
The calculated $c_{5,3}$ approximately matches the hydrodynamic result in Eq.~\eqref{c53} with relative accuracy of 2\%. Given the precision of the interpolating computation, we cannot rule out that, potentially, $c_{5,3} = c_{5,4}$.

While time-consuming in practice, the numerical interpolation can be improved, we claim, to an `arbitrarily good' precision and extended to higher-order terms in the expansion of $\bar q_n$ that will give rise to {\it all} hydrodynamic coefficients $c_{m,j}$. At this point, we could also speculate on further potential underlying symmetry structures of the coefficient triangle \ref{fig:trikotnik} (e.g., it is possible that $c_{m+1,m} = c_{m+1,3}$ for $m\geq 2$, etc.), thereby extending the derived Eq.~\eqref{symmetry_outer} to inner diagonal lines being left-right symmetric about the vertical axis of \ref{fig:trikotnik}, or on there existing other intriguing symmetry structures. However, due to the present lack of conclusive evidence, we defer all such statements until after future detailed analyses have been performed.

\subsection{The reconstruction of rest of the spectrum}
We have so far discussed the reconstruction of the infrared gapless hydrodynamic dispersion relation $\omega_0(q)$ in general $d$. The gradient expansion in Eq.~\eqref{hydro_series} is a convergent series in powers of $\bar q^2$ \cite{Grozdanov:2019kge,Grozdanov:2019uhi}. The convergence properties of the $1/d$ series are for the moment not well understood.\footnote{For some discussion of these issues, see Ref.~\cite{Emparan:2015rva}.} It appears likely to us, however, that, as with the resummed coefficient of $\bar q^4$ (cf.~Eq.~\eqref{hydro_ser_q4_resum}), other coefficients expressed as series in $1/d$ will also converge. Hence, we expect that the dispersion relation \eqref{hydro_series} can be evaluated in any finite integer dimension $d\geq 3$.

Once the series representation of $\omega_0(q)$ in powers of $q^2$ is known, then one can reconstruct the remaining modes' dispersion relations $\omega_{i\geq1}(q)$ that are connected to $\omega_0$ by level-crossings\footnote{For a definition of level-crossing in this context, see Refs.~\cite{Grozdanov:2019uhi,Grozdanov:2021gzh}.} following the procedures developed and demonstrated in Refs.~\cite{Withers:2018srf,Grozdanov:2022npo}. In particular, one needs to re-expand $\omega_0(q)$ around the nearest level-crossing critical point limiting its convergence and use the Puiseux theorem to obtain the expansion of the next mode $\omega_1(q)$ around the same critical point (the second Riemann sheet). Then, after a sequence of analytic continuations and further re-expansions (all which can be done using purely algebraic manipulations), other $\omega_i(q)$ can be derived so long as they do not factor from the full complex spectral curve associated to the correlator (see Refs.~\cite{Grozdanov:2019kge,Grozdanov:2019uhi,Grozdanov:2021gzh}) and are therefore connected to $\omega_0(q)$ via any sequence of level-crossings --- i.e., each is a Riemann sheet of the full surface that can be reconstructed from $\omega_0(q)$. For details of the algorithm, see Ref.~\cite{Grozdanov:2022npo}, where the example of the present setup with $d=3$ (diffusion of a stack of M2 branes \cite{Herzog:2002fn}) was analysed. As long as this reconstruction uncovers all modes in the spectrum of a correlation function, this is sufficient to reconstruct the full correlator meromorphic in $\omega$: the denominator $\prod_{i=0}^\infty \left(\omega - \omega_i(q)\right)$ and also its numerator (see Ref.~\cite{Dodelson:2023vrw}). 

\section{Discussion}
In this Letter, we developed a new interpolation method for uniquely deriving a classical hydrodynamic dispersion relation --- the long-lived excitation in the correlator spectrum of a holographic QFT --- from a discrete set of pole-skipping points in the large-$d$ expansion. Once the dispersion relation is known, it can then be evaluated in any specific dimension $d$ and, as we claim, in some cases, all other modes and the full Green's function can also be reconstructed completely. Conceptually, this reduces at least in principle the computation of a full QFT Green's function to a set of purely algebraic manipulations. We believe that this provides an important step towards answering the question of how much information one requires in order to completely determine a QFT correlation function. Moreover, in the process, we also uncovered an interesting triangular structure underpinning the hydrodynamic coefficients in the (double) gradient and large-$d$ expansions promising significant fertile grounds for future explorations of its mathematical structure. Since our analysis only relies on the analysis of a set of differential equations, these techniques can also be directly used in the analysis of QNM spectra of black holes, analysis of complex (or algebraic) spectral curves, and other mathematical problems that pertain to the spectral analysis of linear operators. 

Many open questions remain for the future. It should be understood whether the powerful techniques developed in Ref.~\cite{Aminov:2023jve} can be applied to the present analysis. Moreover, it would be interesting to show how precisely a similar interpolation works for spectra with sound modes or potentially even for spectra without any hydrodynamic modes. How does the reconstruction work for spectra in which the diffusive mode collides with a gapped mode for real $q$ in the sense of Ref.~\cite{Grozdanov:2018fic}? One may also ask whether an interpolation not using the pole-skipping points might prove to be useful. Finally, given the promise that this procedure has for reducing the complexities in computing various QFT correlators in the strongly-coupled large-$N$ limit, it is natural to wonder whether an analogous procedure can be devised once we start including the inverse coupling constant corrections (as e.g.~in Ref.~\cite{Grozdanov:2016vgg}) and the significantly more difficult `$1/N$' corrections in the holographic description, or even in non-holographic QFTs without the help of gravity.

\begin{acknowledgments}
We would like to thank Borut Bajc, Richard Davison, Roberto Emparan, Giorgio Frangi, Alexander Soloviev and Mile Vrbica for illuminating discussions. The work of S.G. was supported by the STFC Ernest Rutherford Fellowship ST/T00388X/1, the research programme P1-0402 and the project N1-0245 of Slovenian Research Agency (ARIS). T.L. is supported by the research programme P1-0402 of Slovenian Research Agency (ARIS). J.F.P. is supported by the `Atracci\'on de Talento' program (Comunidad de Madrid) grant 2020-T1/TIC-20495, by the Spanish Research Agency via grants CEX2020-001007-S and PID2021-123017NB-I00, funded by MCIN/AEI/10.13039/501100011033, and by ERDF A way of making Europe.
\end{acknowledgments}

\bibliographystyle{apsrev4-1}
\bibliography{Genbib}{}

\onecolumngrid
\appendix
\section{The determinant method for computing the pole-skipping points}\label{app:Determinant}

A conceptually clear and systematic way to obtain the pole-skipping points  $(\omega_n, q_n)$ is to utilise the {\em determinant method} of Ref.~\cite{Blake:2019otz}. 
Here, we briefly review this procedure and write a closed-form expression for the relevant determinant for all pole-skipping points that follow from the differential equation \eqref{EoM} in the large-$d$ expansion. The pole-skipping points are determined from Eq.~\eqref{EoM} by expanding the equation of motion in a power series around the horizon and demanding the solution has two independent parameters. This is synonymous with the `indeterminacy' of pole-skipping (see e.g. the discussion in Ref.~\cite{Grozdanov:2023txs}). It can be easily seen that the frequencies are given by $\omega_n = -2\pi T i n$, with $n=0,1,2,\ldots$, while the momenta $q_n$ are given as solutions to the equation 
\begin{equation}
    \det(M^{(n)}) = 0,
    \label{detEquation}
\end{equation}
where $M^{(n)}$ is an $n\times n$ matrix following from the first $n$ equations of the expansion of the equation \eqref{EoM} around the horizon.

The matrix $M^{(n)}$ is of the form
\begin{equation}
    M^{(n)} = \begin{pmatrix}
        \gamma_0 & \beta_0 & 0 & 0 & \dots \\
        \gamma_1 & \beta_1 + \gamma_0 & 2\alpha_1+2\beta_0 & 0 & \dots \\
        \gamma_2 & \beta_2 + \gamma_1 & 2\alpha_2+2\beta_1+\gamma_0 & 6\alpha_1+3\beta_0 & \dots \\
        \gamma_3 & \beta_3 + \gamma_2 & 2\alpha_3+2\beta_2+\gamma_1 & 6\alpha_1+3\beta_0+\gamma_0 & \dots \\
        \vdots & \vdots & \vdots & \vdots & \ddots
    \end{pmatrix} , \label{Moblika}
\end{equation}
where the matrix elements are given by the following expression:
\begin{equation}
    M^{(n)}_{ij} = (j - 1) (j - 2) \alpha_{i-j+2} + (j - 1) \beta_{i-j+1} + \gamma_{i-j},
\end{equation}
with the coefficients $\alpha_j$, $\beta_j$ and $\gamma_j$ given by
\begin{align}
    \alpha_j &= \frac{(-1)^j n d^2 }{j!}\left[2 \left(2 b^2+n\right) (1-2 d)_{j-1}-\left(4 b^2+n\right) (1-d)_{j-1}\right], \\
    \beta_j &= (-1)^j n d \, \Biggr[ \frac{(d-2) \left(j \left(4 b^2 (d+1)+n\right) (3-d)_{j-2}+\left(2 b^2 (d (n-2)-2)-n\right)
   (3-d)_{j-1}\right)}{j!} \nonumber \\
   &\phantom{(-1)^j n d \Bigg(\quad} -\frac{(-1)^j (j+1) \left(2 b^2+n\right) \Gamma (2 d-1) (d (-2 d (n-1)+j n+n+1)-1)}{\Gamma (j+2) \Gamma (2d-j)}-2 b^2 (d+1)\Biggr] , \\
   \gamma_j &= \frac{(-1)^j n^2 d^2}{j!}\Big[ (d-2) (d-1) j \left(2 b^2+n\right) (5-2 d)_{j-2}-(d-2) \left(b^2 d+d-1\right) \left(2 b^2+n\right) (5-2d)_{j-1} \nonumber \\
   &\phantom{\frac{1}{j!}d^2 (-1)^j n^2 \Big(}+b^2 \left((2 d-1) j (4-d)_{j-1}-\left(\left(b^2+2\right) d-1\right) (4-d)_j\right)\Big].
\end{align}
In stating the above expressions, we used the momentum $\bar{q} = b \sqrt{nd/2}$. Moreover, $(x)_n$ is the Pochhammer symbol and as can be inferred from Eq.~\eqref{Moblika}, the two coefficients $\beta_j$ and $\gamma_j$ are defined for $j\geq 0$, while $\alpha_j$ is defined for $j\geq 1$.

Assuming that $b$ is the rescaled pole-skipping momentum $b_n$ (cf.~Eq.~\eqref{qbar_def}): 
\begin{equation}
    b_n=1 + \sum_{m=1}^\infty \frac{b_{n,m}}{d^m},
\end{equation}
we find the components of the matrix $M^{(n)}$ to have the following form: 
\begin{equation}
    M^{(n)}_{ij} = \sum_{r=-(i-j+3)}^\infty \frac{m_{ij,r}}{d^r}.
\end{equation}
Here, we will not state the coefficients $m_{ij,r}$ explicitly as they are too complicated and do not provide much intuition into the computation. What is relevant, however, is that they can be easily computed analytically and that they contain the coefficients $b_{n,m}$ and their convolutions. The values of $b_{n,m}$ are thus determined by solving the equation~\eqref{detEquation}, where the determinant can now be more explicitly written in the expansion in powers of $1/d$ as
\begin{equation}
\det M^{(n)} = \sum_{k=-3n}^\infty \frac{D_k}{d^k},
\end{equation}
where
\begin{equation}
D_k = \sum_{i_1,\dots,i_n=1}^n \varepsilon_{i_1,\dots,i_n} D_{i_1,\dots,i_n,k}
\end{equation}
and
\begin{equation}
D_{i_1,\dots,i_n,k} = \sum_{r_{n-1}=-3(n-1)}^{k + (n-i_n+3)} \sum_{r_{n-2}=-(n-2-i_{n-2}+3)}^{r_{n-1}+(n-1-i_{n-1}+3)} \cdots \sum_{r_1=-(1-i_1+3)}^{r_2+(2-i_2+3)} m_{1 i_1, r_1} m_{2 i_2, r_2-r_1} \dots m_{(n-1) i_{n-1}, r_{n-1}-r_{n-2}} m_{n i_n, k-r_{n-1}}.
\end{equation}

In practice, we solve for the coefficients $b_{n,m}$ for a specific $m=M$. We do this by choosing a maximal $n=n_{\text{max}}$ and then solve Eq.~\eqref{detEquation} for different $n=1,\dots, n_\text{max}$, only expanding the coefficients of the matrix $M^{(n)}$ in $1/d$ to power $M$. In this way, we obtain a list of the first few values of $b_{n,M}$ that determine the coefficients $c_{m,m+M-1}$. In our case, they follow either from an analytic analysis or from a Padé approximant that we fit to $b_{n,M}$ and use to compute the derivatives of $b_{n,M}$ at $n=0$.

\section{Obtaining  $b_{n,m}$ from the quasinormal mode calculation}\label{app:FullB}

We start by writing the differential equation \eqref{EoM} for the gauge invariant mode $Z$ in terms of a new coordinate $x=r^d$. The horizon is located at $r=1$ and the boundary at $x \to \infty$. The ordinary differential equation (ODE) now reads 
\begin{equation}
    \partial^2_x Z(x) + P(x) \partial_x Z(x) + Q(x)Z(x) = 0,
\end{equation}
where
\begin{align}
P(x, \omega, \bar{q}, d) &= \frac{1}{x-1} \left( 2-\frac{2}{x} -\frac{2i\omega x^{-1/d}}{d} + \frac{\omega^2}{\omega^2 x - \bar{q}^2 (x-1)d} \right), \\
Q(x, \omega, \bar{q}, d) &= \frac{x^{-1-2/d}}{(x-1)d^2} \left( \frac{i\omega x^{1/d} \left[ ((x-2)d-(x-1))\bar{q}^2 d - \omega^2x(d-1) \right]}{\omega^2x -(x-1)\bar{q}^2d} - \bar{q}^2 d \right).
\end{align}
Recall that $\bar{q} \equiv q/\sqrt{d}$.

Since we are interested in finding the values of the hydrodynamic dispersion relation at the pole-skipping points, we set $\omega = - 2 \pi T i n = - i d n / 2$ and expand the differential equation term-by-term in a power series of $1/d$. The gauge-invariant mode $Z(x)$, functions $P(x)$, $Q(x)$, and the pole-skipping momentum $\bar{q}_n$ take the form
\begin{equation}
Z(x) = \sum_{m=0}^\infty \frac{Z_m(x)}{d^m}, \quad P(x)=\sum_{m=0}^\infty \frac{p_m(x)}{d^m}, \quad
Q(x)=\sum_{m=0}^\infty \frac{q_m(x)}{d^m}, \quad
\bar{q}_n = \sqrt{\frac{nd}{2}}\left(1 + \sum_{m=1}^\infty \frac{b_{n,m}}{d^m} \right), 
\end{equation}
which turns the ODE into an infinite system of non-homogeneous equations of the form
\begin{equation}
    \partial_x^2 Z_m + p_0 \partial_x Z_m + q_0 Z_m + F_m = 0,
\end{equation}
for $m=0,1,2,\ldots$. The $M$-th non-homogeneous part $F_{M}$ depends on the solutions to all of the equations with $m < M$:  
\begin{equation}
    F_m = \sum_{k=1}^m \left(\partial_x^2 Z_{m-k} + p_k \partial_x Z_{m-k} + q_k Z_{m-k}\right),
\end{equation}
while the homogeneous part of the equation is independent of the level $m$. 

We start by solving the zeroth-order equation in $1/d$:
\begin{equation}
\partial_x^2 Z_0 + p_0 \partial_x Z_0 + q_0 Z_0 = 0,
\end{equation}
that has two independent solutions:
\begin{align}
z_1(x) &= \frac{1}{x} \label{indep_z1} \\
z_2(x) &= \frac{(n+2)x-1}{(n+1)x}(x-1)^n . \label{indep_z2}
\end{align}
Imposing the ingoing boundary conditions at the horizon (which in the Eddington-Finklestein coordinate translate to imposing regularity) and the vanishing Dirichlet boundary condition at the boundary (the standard QNM condition),
\begin{equation}
\begin{aligned}
\lim_{x\to 1} Z(x) &= 1 \\ 
\lim_{x\to \infty} Z(x) &= 0,
\label{Zbcs}
\end{aligned}
\end{equation}
implies the QNM solution
\begin{equation}
Z_0(x) = \frac{1}{x}.
\end{equation}

Each subsequent $Z_m$ then depends on all the previous solutions $Z_0, \dots, Z_{m-1}$. In general, it is given by the expression
\begin{equation}
Z_m(x) = c_1z_1(x) + c_2z_2(x) + z_1(x) \int_1^x \frac{z_2(y) F_m(y)}{W(y)} dy - z_2(x) \int_1^x \frac{z_1(y) F_m(y)}{W(y)} dy,
\label{eq:ZmIntegral}
\end{equation}
where $W(x)=z_1(x) z_2'(x) - z_2(x) z_1'(x)$ is the Wronskian of the two homogeneous solutions $z_1$ and $z_2$ in Eqs.~\eqref{indep_z1} and \eqref{indep_z2}, and the coefficients $c_1$ and $c_2$ are determined by the boundary conditions in Eq.~\eqref{Zbcs}.

The second of the two integrands above is of the form
\begin{equation}\label{log-hor}
    \frac{z_1(x)F_m(x)}{W(x)} = \frac{n F_m(x)}{4 x^2} \frac{1}{(x-1)^n} ,
\end{equation}
which implies that the removal of the logarithmic divergence at the horizon translates into the result for $b_{n,m}$. Explicitly, to find each $b_{n,m}$, we expand Eq.~\eqref{log-hor} around $x=1$ as 
\begin{equation}
\frac{n F_m(x)}{4 x^2} = \sum_{k=0}^\infty f_{m,k} (x-1)^k.
\label{Fmexpansion}
\end{equation}
The residue of the discussed integrand in Eq.~\eqref{eq:ZmIntegral} is set by the $(n-1)$-th coefficient $f_{m,n-1}$, which controls the term that causes the logarithmic divergence in $Z_m(x)$. Hence, by solving the equation $f_{m,n-1}=0$ at each $m$, we determine $b_{n,m}$.

At $m=1$, we find that
\begin{equation}
    b_{n,1} = -\frac{1}{2} H_n,
\end{equation}
which then by using Eq.~\eqref{cmm} allows us to deduce all coefficients $c_{m,m}$ stated in Eq.~\eqref{symmetry_outer}.

At $m=2$, the calculation is significantly more involved. We first solve for $Z_1$, which we then insert into $F_2$ and use the expansion \eqref{Fmexpansion} to obtain $b_{n,2}$. We find
\begin{align}
    Z_1(x) &= \frac{2H_n(n\ln(x) + (n+2)(x-1))+\ln(x)\left(n\ln(x) + 2(n+2)x-2\right)}{4(n+1)x} \nonumber \\
    &- \frac{(n+2)x - 1}{2n(n+1)x} \left(\frac{x-1}{x}\right)^n \left[ \left(n H_n+ n \ln(x) -1\right)  {}_2F_1\left(n,n;n+1;\frac{1}{x}\right) + {}_3F_2\left(n,n,n;n+1,n+1;\frac{1}{x}\right)\right],
\end{align}
where ${}_2F_1$ and ${}_3F_2$ are hypergeometric functions. Finally, we state here the expression for $b_{n,2}$ in form of a double sum, which can be used in the numerical interpolation to compute $c_{m+1,m}$. The result is
\begin{align}
    b_{n,2} &= a_n + b_n H_n + c_n H_n^{(2)} + d_n H_n H_n^{(2)} + e_n H_n^2 + \sum_{k=1}^{n-3} \left( j_{n,k} + k_{n,k} H_n + l_{n,k} H_k + m_{n,k} H_k^{(2)} \right) \nonumber \\
    &+ \frac{n}{8(n+1)} \left[\sum_{k=1}^{n-3} \frac{1}{k+1} \left(\frac{n+2}{n}\right)^k\right]\left[\sum_{k=1}^{n-3} \frac{1}{k+1} \left(\frac{n}{n+2}\right)^k\right] - \frac{3n}{8(n+1)} \sum_{k=2}^{n-3} \frac{1}{k+1} \left(\frac{n}{n+2}\right)^k \sum_{j=0}^{k-2} \frac{1}{j+1} \left(\frac{n+2}{n}\right)^j \nonumber \\
    &+ \frac{1}{2(n+1)} \sum_{k=2}^{n-3} H_k \left(\frac{n}{n+2}\right)^k \sum_{j=0}^{k-2} \frac{1}{j+1} \left(\frac{n+2}{n}\right)^j + \frac{n+2}{4n(n+1)} \sum_{k=2}^{n-3} \frac{1}{k-1} \left(\frac{n}{n+2}\right)^{k-1} \sum_{j=k}^{n-3} \frac{1}{j+1} \left(\frac{n+2}{n}\right)^j \nonumber \\
    &+ \frac{n+2}{4n(n+1)} \sum_{k=2}^{n-3} \frac{1}{k-1} \left(\frac{n}{n+2}\right)^{k-1} \sum_{j=k}^{n-3} H_j \left(\frac{n+2}{n}\right)^j + \frac{3n^2}{8(n+1)(n+2)} \sum_{k=1}^{n-3} \frac{1}{k} \sum_{j=k+1}^{n-3} \frac{1}{j} \left(\frac{n+2}{n}\right)^{n-j},
\end{align}
where $H^{(m)}_n = \sum_{k=1}^n 1 / k^m $ is the generalised Harmonic number. The remaining expressions that appear in $b_{n,2}$ are 
\begin{align}
    a_n &= \frac{1-2 n}{4 \left(n^2-1\right)} \left(\frac{n}{n+2}\right)^{n-3} + \frac{12-n (2 (n-3) (n-1) n+11)}{8 n^2 \left(n^2-4\right)^2 \left(n^2-1\right)} \left(\frac{n+2}{n}\right)^n \nonumber \\
    &\phantom{=}- \frac{n^9-17 n^8+77 n^7-103 n^6-44 n^5+4 n^4+378 n^3-544 n^2+392 n-96}{8 (n-2)^2 (n-1)^3 n^2 (n+1) (n+2)}, \\
    b_n &= \frac{n^7-10 n^6+21 n^5-32 n^4-28 n^3+72 n^2-96 n+96}{16 (n-2)^2 (n-1)^2 n (n+1) (n+2)} \nonumber \\
    &\phantom{=}+\frac{(5 n+6)}{8
   (n+1)}\left(\frac{n}{n+2}\right)^{n-2} - \frac{3 (n-2)}{16 (n+1)} \left(\frac{n+2}{n}\right)^{n-2} ,\\
   c_n &= -\frac{n \left(6 n^2-16 n+9\right)}{8 (n+1) \left(n^2-3 n+2\right)}, \\
   d_n &= -\frac{n}{4 n+4}, \\
   e_n &= \frac{5 n^3-8 n^2-10 n+10}{8 \left(n^3-2 n^2-n+2\right)} ,
\end{align}
and
\begin{align}
   j_{n,k} &= \Bigg(\frac{(k (k (5 k+24)+23)+8) n^2-2 (3 k+1) (k+1)^2 n+4 k (k+1)^2}{8 k (k+1)^2 (n-2) (n-1) n (n+1) (k-n+1)} \nonumber \\
   &\phantom{=} \phantom{\Bigg(} +\frac{-k (k+2) n^5+(k (k (3 k+10)+11)+3) n^4+(-k (k+2) (7 k+12)-9) n^3}{8 k (k+1)^2 (n-2) (n-1) n (n+1) (k-n+1)} \Bigg) \left(\frac{n+2}{n}\right)^k \nonumber \\
   &\phantom{=}+ \frac{k (n (n (n+3)+6)-4)+n (6-(n-3) n (n+1))-4}{8 (k+1) \left(n^2-1\right) (k-n+1)} \left(\frac{n}{n+2}\right)^k \nonumber \\
   &\phantom{=}+ \frac{2 k^2 n^2-5 k^2 n-2 k n^3+8 k n^2-9 k n+n^3-3 n^2+2 n}{8 k^3 (k+1) (n-2) (n-1) (n+1)}, \\
   k_{n,k} &= \frac{2 k^2 (k+1)^2+(k (k+4)+1) n^2-2 k (k+1)^2 n-n^3}{8 k (k+1) n (n+1) (k-n+1)} \left(\frac{n+2}{n}\right)^k \nonumber \\
   &\phantom{=}+ \frac{-2 k (n+1) (n+6)+n (n (2 n+7)-18)-24}{8 (n+1) (n+2) (-k+n-2) (-k+n-1)} \left(\frac{n}{n+2}\right)^k, \\
   l_{n,k} &= \frac{n \left(4 k (n+2)-3 n^2+4 n+8\right)}{8 (k+1) (n+1) (n+2)^2} \left(\frac{n+2}{n}\right)^{n-k} -\frac{(n-2 k)}{8 k (k+1) (n+1)} \left(\frac{n+2}{n}\right)^k \nonumber \\
   &\phantom{=}+ \frac{4 (k (k+2)-2) (k (k+11)+6) n^2-8 k (k+2) (4 k-1) (k+1) n+16 k (k+2) (k+1)^2}{8 k^2 (k+1)^2 (k+2) n \left(n^4-5 n^2+4\right)} \nonumber \\
   &\phantom{=}+ \frac{(k (k (5 (k-1) k+4)+102)+44) n^4+2 (k (k (k (17 k+33)-32)-32)+8) n^3}{8 k^2 (k+1)^2 (k+2) n \left(n^4-5 n^2+4\right)} \nonumber \\
   &\phantom{=}+ \frac{4 \left(k^3-5 k-2\right) n^6+(k (10-3 k (k (3 k+7)-2))-4) n^5}{8 k^2 (k+1)^2 (k+2) n \left(n^4-5 n^2+4\right)}, \\
   m_{n,k} &= \frac{(k-1) n}{4 k (k+2) (n+1)}.
\end{align}
It is plausible that a further simplification of this expression could lead to a closed-form analytic expression for $c_{m+1,m}$. Note also that in order to avoid unnecessary divergences, a certain amount of care is needed when evaluating these expressions for small values of $n$. 

\twocolumngrid

\end{document}